\documentstyle[12pt,myepsfig]{article}
\begin{document}

\begin{center}

{\LARGE\bf Beaming Effects in Gamma-Ray Bursts}

\vspace{5mm}

{\large\bf Huang Y.F.$^1$, Lu T.$^1$, Dai Z.G.$^1$ and Cheng K.S.$^2$}

\vspace{2mm}

{\small\bf $^1$ Department of Astronomy, Nanjing University,
        Nanjing 210093, China; hyf@nju.edu.cn} 

{\small\bf $^2$ Department of Physics, the University of Hong Kong,
                     Hong Kong, China}

\end{center}

\begin{abstract}

Based on a refined generic dynamical model, we investigate
afterglows from jetted gamma-ray burst (GRB) remnants numerically. 
In the relativistic phase, the light curve break could marginally 
be seen. However, an obvious break does
exist at the transition from the relativistic phase to the non-relativistic
phase, which typically occurs at time 10 to 30 days. It is very interesting
that the break is affected by many parameters, especially by the electron energy
fraction ($\xi_{\rm e}$), and the magnetic energy fraction ($\xi_{\rm B}^2$).
Implication of orphan afterglow surveys on GRB beaming is investigated. 
The possible existence of a kind of cylindrical jets is also discussed.

\end{abstract}

\vspace{0.5cm}

\section{Introduction}

Researches on afterglows from long gamma-ray bursts (GRBs) have shown 
that they are of cosmological origin. 
The standard fireball model,
which incorporates internal shocks to explain the main 
bursts and external shocks to account for 
afterglows,
becomes the most popular model (for recent reviews, see van 
Paradijs et al. 2000). Some GRBs 
localized by BeppoSAX satellite
have implied isotropic energy release of more than $10^{54}$ ergs,
leading many theorists to deduce that GRB radiation must be
highly collimated (Castro-Tirado et al. 1999; Huang 2000;
Halpern et al. 2000; Dai \& Gou 2001; Dai \& Cheng 2001; 
Gou et al. 2001; Ramirez-Ruiz \& Lloyd-Ronning 2002; 
Zhang \& M\'{e}sz\'{a}ros 2002).

To differentiate a jet from an isotropic fireball, we must
resort to the afterglow light curves. When the bulk Lorentz factor of
a jet drops to $\gamma < 1 / \theta$, with $\theta$ the half opening
angle, the edge of the jet becomes visible, the light curve will
steepen by $t^{-3/4}$. This is called the edge effect (M\'{e}sz\'{a}ros
\& Rees 1999; Panaitescu \& M\'{e}sz\'{a}ros 1999; Kulkarni et al. 1999). 
In addition, the lateral expansion of a relativistic
jet will make the break even more precipitous. 
So it is generally believed that afterglows from jetted GRBs are
characterized by an obvious break in the light curve at
{\em the relativistic stage}.

In this talk, we use our refined dynamical model to study the jet
effect on the afterglow light curves. The possible existence of
cylindrical jets is also discussed.

  \begin{figure}[htb]
  \begin{center}
  \leavevmode
  \centerline{ \hbox{ \hbox{} \hspace{0.5in}
  \epsfig{figure=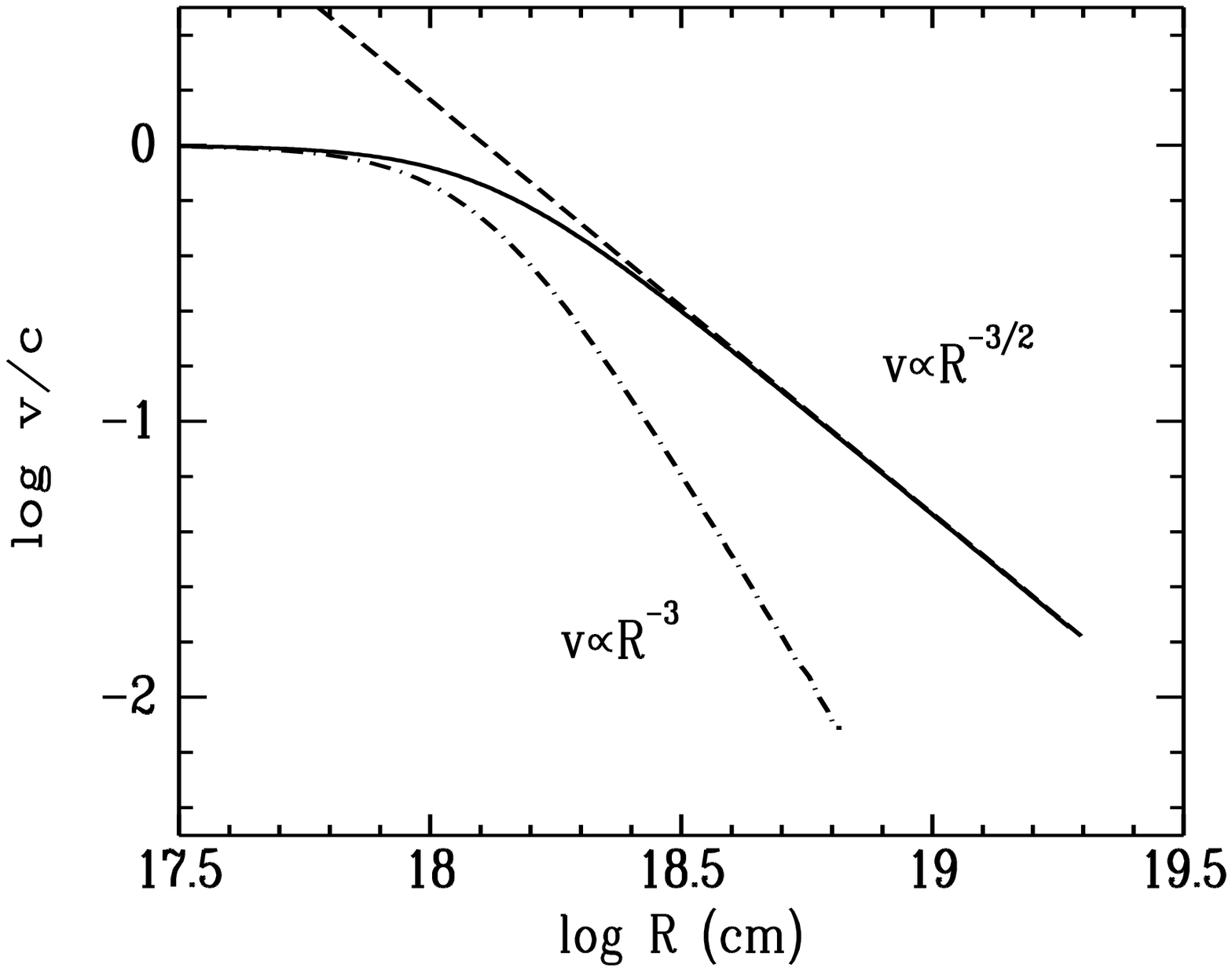,width=2.38in,height=1.2in,angle=-90,
bbllx=200pt, bblly=270pt, bburx=450pt, bbury=680pt}
  \hspace{0.5in}
  \epsfig{figure=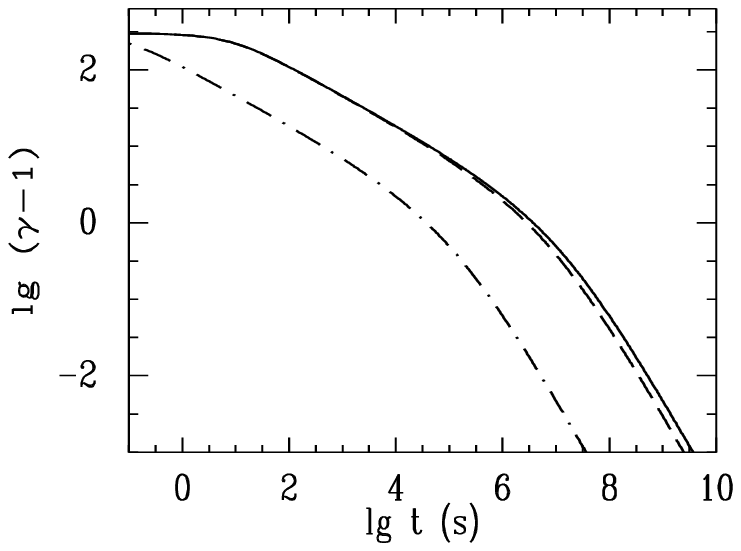,width=2.38in,height=1.2in,angle=-90,
bbllx=230pt, bblly=350pt, bburx=330pt, bbury=530pt}
  }}
  \begin{flushright}
  \parbox[t]{2.9in} { \caption {  Velocity vs. radius for an isotropic adiabatic 
fireball (Huang 2000). The dashed line is the familiar Sedov solution in 
the Newtonian phase. The dash-dotted line is drawn according to Eq.~(1),
which differs from the dashed line markedly. The solid line
corresponds to our refined model (i.e., Eq.~(2)), which is
consistent with the Sedov solution (Huang 2000).   }} \ \hspace{.2in} \
  \parbox[t]{2.9in} { \caption { Evolution of $\gamma$. The solid line
corresponds to a jet with ``standard'' parameters. Other lines are drawn
with only one parameter altered: the dashed line corresponds 
to $\theta_0 = 0.1$, and the dash-dotted line corresponds 
to $n = 10^6$ cm$^{-3}$ (Huang et al. 2000c). }}
  \end{flushright}
  \end{center}
  \end{figure}

\section{Dynamics}

The importance of the non-relativistic phase of fireball expansion has
been stressed by Huang et al. (1998a, b). 
In the literature, it is
generally believed that the following equation can depict the evolution
of GRB remnants,
\begin{equation}
\label{dgdm1}
\frac{d \gamma}{d m} = - \frac{\gamma^2 - 1}{M},
\end{equation}
where $m$ is the rest mass of the swept-up medium, $M$ is the total mass
in the co-moving frame, including internal energy. However, Huang
et al. (1999) pointed out that during the non-relativistic phase
of an adiabatic expansion, Eq.~(1) cannot reproduce the familiar Sedov
solution. This is clearly shown in Fig.~1.

Huang et al. (1999) have proposed a refined equation,
\begin{equation}
\label{dgdm2}
\frac{d \gamma}{d m} = - \frac{\gamma^2 - 1}
       {M_{\rm ej} + \epsilon m + 2 ( 1 - \epsilon) \gamma m}, 
\end{equation}
where $M_{\rm ej}$ is the initial baryon mass ejected from the GRB
central engine, and $\epsilon$ is the radiative efficiency. For an
adiabatic fireball, $\epsilon = 0$; and for a highly radiative one,
$\epsilon = 1$. Huang et al. (1999) have shown that Eq.~(2) is 
correct for both radiative and adiabatic fireballs, and in both
ultra-relativistic and non-relativistic phases (c.f. Fig.~1).

Using the refined dynamical model, the evolution of the beamed ejecta
can be described by (Huang et al. 2000a, b, c):
\begin{equation}
\label{drdt1}
\frac{d R}{d t} = \beta c \gamma (\gamma + \sqrt{\gamma^2 - 1}),
\end{equation}
\begin{equation}
\label{dmdr2}
\frac{d m}{d R} = 2 \pi R^2 (1 - \cos \theta) n m_{\rm p},
\end{equation}
\begin{equation}
\label{dthdt3}
\frac{d \theta}{d t} = \frac{c_{\rm s} (\gamma + \sqrt{\gamma^2 - 1})}{R},
\end{equation}
\begin{equation}
\label{dgdm4}
\frac{d \gamma}{d m} = - \frac{\gamma^2 - 1}
       {M_{\rm ej} + \epsilon m + 2 ( 1 - \epsilon) \gamma m}, 
\end{equation}
\begin{equation}
\label{cs5}
c_{\rm s}^2 = \hat{\gamma} (\hat{\gamma} - 1) (\gamma - 1) 
	      \frac{1}{1 + \hat{\gamma}(\gamma - 1)} c^2 , 
\end{equation}
where $m_{\rm p}$ is the proton mass, $c_{\rm s}$ is the co-moving 
sound speed, $\hat{\gamma} \approx (4 \gamma + 1)/(3 \gamma)$ is the 
adiabatic index, and $c$ is the speed of light. Below we will  
consider only adiabatic jets, for which $\epsilon \equiv 0$.  

\section{Beaming Effects}

A strong blastwave will be generated due to the interaction of the 
jet and the ISM. Synchrotron radiation from the shock accelerated 
ISM electrons gives birth to afterglows. As usual we assume that 
the magnetic energy density in the co-moving frame is a 
fraction $\xi_{\rm B}^2$ of the total thermal energy density 
($B'^2 / 8 \pi = \xi_{\rm B}^2  e'$), and that electrons carry 
a fraction $\xi_{\rm e}$ of the proton energy. This means that the 
minimum Lorentz factor of the random motion of electrons in 
the co-moving frame is 
$\gamma_{\rm e,min} = \xi_{\rm e} (\gamma - 1) 
		     m_{\rm p} (p - 2) / [m_{\rm e} (p - 1)] + 1$, 
where $p$ is the index characterizing the power law energy distribution 
of electrons, and $m_{\rm e}$ is the electron mass. 

For convenience, let us define the following initial values or parameters 
as a set of ``standard'' parameters: initial energy per solid 
angle $E_0 / \Omega_0 = 10^{54}$ ergs/$4 \pi$, $\gamma_0 = 300$,
$n = 1$ cm$^{-3}$, $\xi_{\rm B}^2 = 0.01$, $p = 2.5$, 
$\xi_{\rm e} =0.1$, $\theta_0 = 0.2, \theta_{\rm obs}=0$, 
$D_{\rm L} = 10^6$ kpc, where $\theta_{\rm obs}$ is 
the angle between the line of sight and the jet axis, and $D_{\rm L}$ 
is the luminosity distance. 

We have followed the evolution of jetted GRB remnants numerically and 
calculated their afterglows (Huang et al. 2000a, c).
Fig. 2 shows the evolution of $\gamma$
for some exemplary jets. We see that the ejecta will cease 
to be highly relativistic at time $t \sim 10^5$ --- $10^6$ s. 
This gives strong support to our previous argument that we should
be careful in discussing the fireball evolution under the simple 
assumption of ultra-relativistic limit (Huang et al. 1998a, b, 1999,
2000a, b, c).

  \begin{figure}[htb]
  \begin{center}
  \leavevmode
  \centerline{ \hbox{ \hbox{} \hspace{0.5in}
  \epsfig{figure=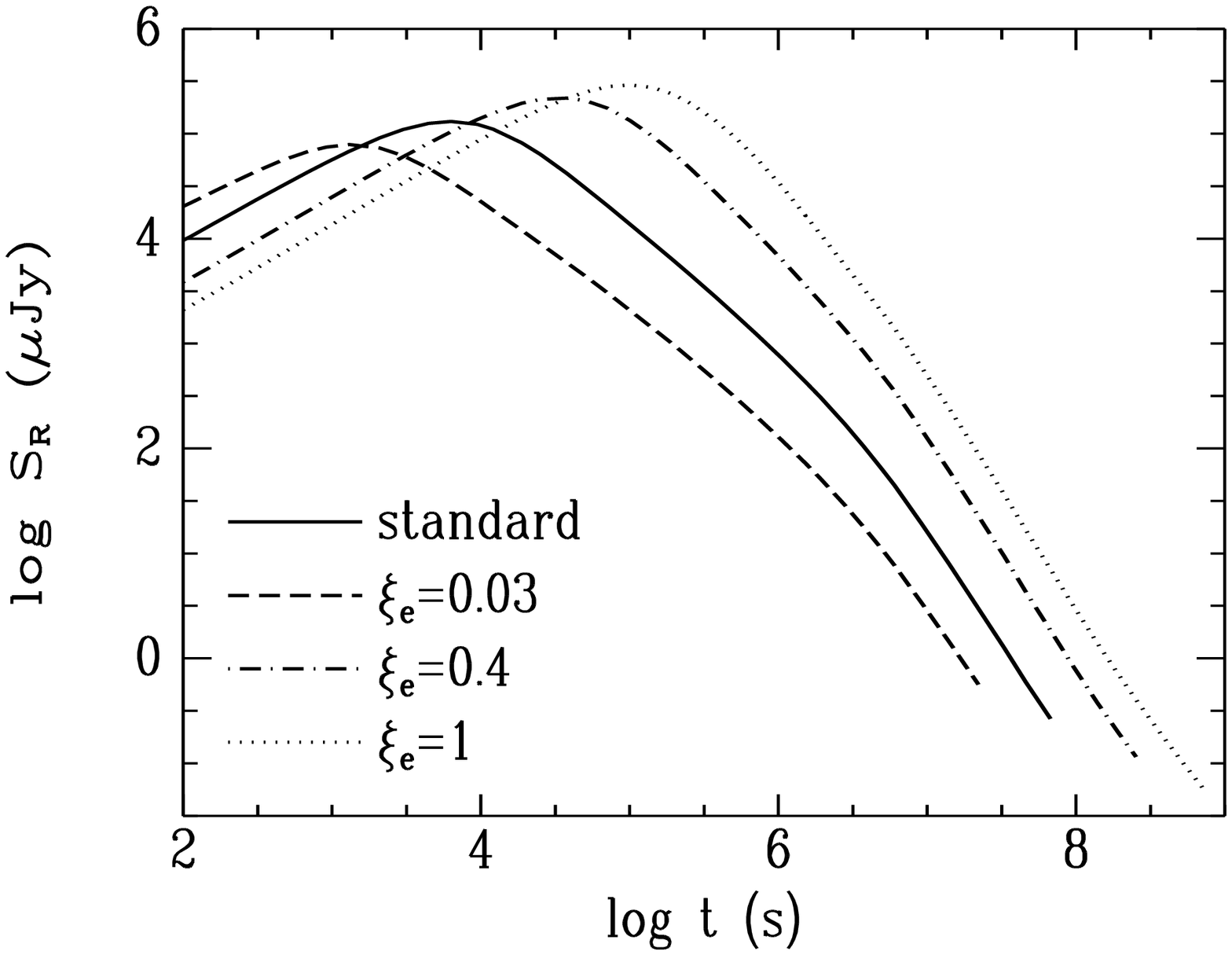,width=2.38in,height=1.2in,angle=-90,
bbllx=195pt, bblly=210pt, bburx=465pt, bbury=690pt}
  \hspace{0.5in}
  \epsfig{figure=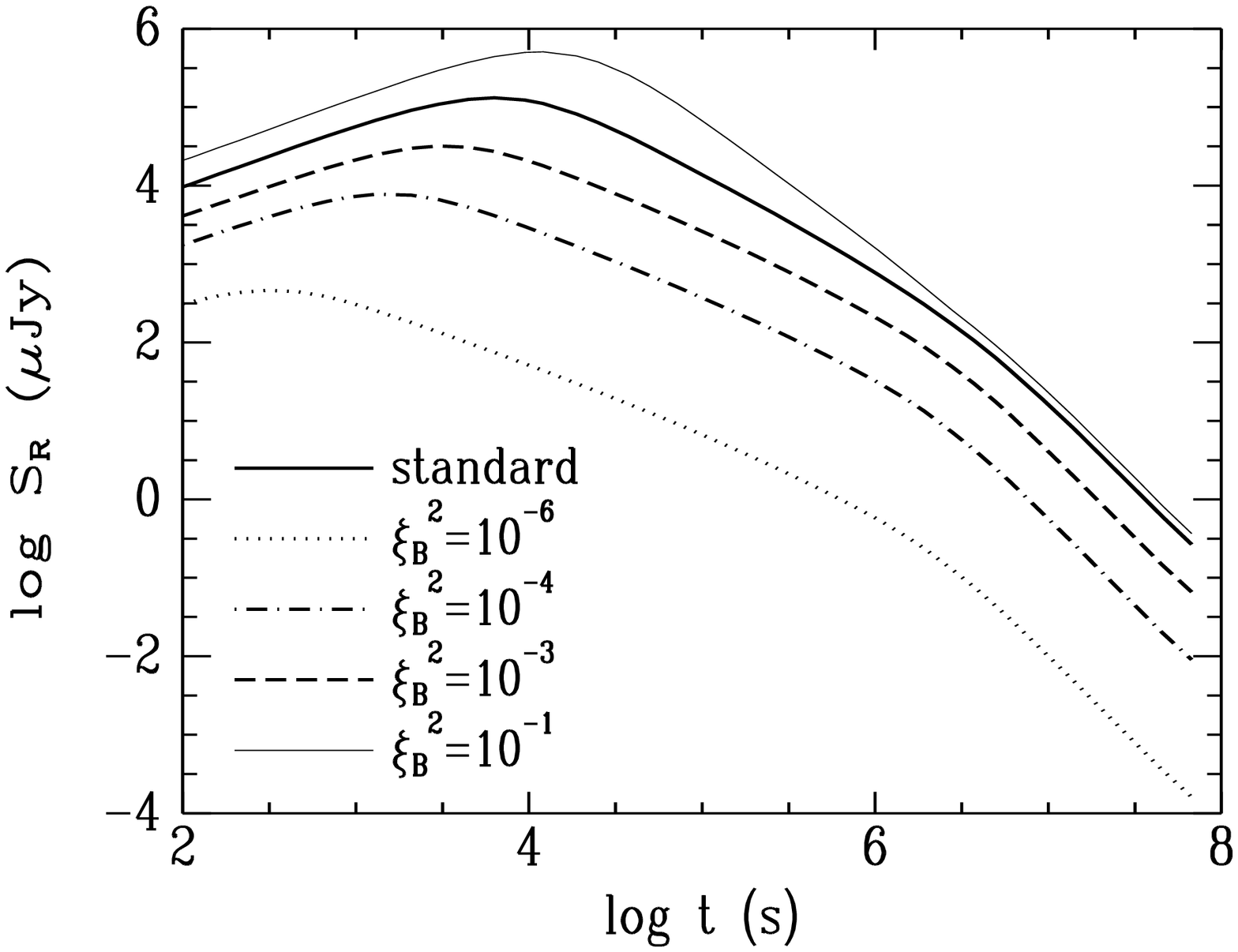,width=2.38in,height=1.2in,angle=-90,
bbllx=195pt, bblly=180pt, bburx=465pt, bbury=660pt}
  }}
  \begin{flushright}
  \parbox[t]{2.9in} { \caption {  The effect of 
$\xi_{\rm e}$ on the R-band light curve.  
The solid line corresponds to a jet with ``standard'' parameters.
Other lines are drawn with only $\xi_{\rm e}$ altered 
(Huang et al. 2000c).  }} \ \hspace{.2in} \
  \parbox[t]{2.9in} { \caption { The effect of 
$\xi_{\rm B}$ on the R-band light curve (Huang et al. 2000c). }}
  \end{flushright}
  \end{center}
  \end{figure}

Fig. 3 illustrates the effect of $\xi_{\rm e}$ on the optical 
(R-band) light curves. In no case could we observe the theoretically 
predicted light curve steepening (with the break point determined 
by $\gamma \sim 1/\theta$) during {\em the relativistic stage itself}.
The reason is: at
time of $\gamma \sim 1/\theta$, the jet is already in its mildly 
relativistic phase and it will become non-relativistic soon after that, 
so the break due to the edge effect and the lateral expansion
effect does not have time to emerge during the relativistic 
phase (Huang et al. 2000a, c).
Further more, since $\gamma$ is no longer much larger
than 1, conventional theoretical analyses (under the assumption 
of $\gamma \gg 1$) are not proper. However, when $\xi_{\rm e}$ is 
small, an obvious break does appear in the light curve, but it is clearly 
due to the relativistic-Newtonian transition.  The simulations by 
many other authors do not reveal such breaks, because their models 
are not appropriate for Newtonian expansion. When $\xi_{\rm e}$ is 
large, the break disappears. This is not difficult to understand.
According to the analysis in the ultra-relativistic limit, the 
time that the light curve peaks scales as 
$t_{\rm m} \propto \xi_{\rm e}^{4/3} (\xi_{\rm B}^2)^{1/3}$.
Fig. 3 shows this trend qualitatively. 
In the case of $\xi_{\rm e} = 1.0$, $t_{\rm m}$ is as large 
as $\sim 10^5$ s, then we can not see the initial power law 
decay (with timing index $\alpha \sim 1.1$) in the relativistic 
phase, it is hidden by the peak.
So the break disappears. We should also note that in 
all cases, light curves during the non-relativistic phase 
are characterized by quick decays, with $\alpha \geq 2.1$. 
This is quite different from isotropic fireballs, whose light curves 
steepen only slightly after entering the Newtonian phase 
(i.e., $\alpha \sim 1.3$). 

Fig. 4 illustrates the effect of $\xi_{\rm B}^2$ on the optical 
light curves. Interestingly but not surprisingly, we see that 
$\xi_{\rm B}^2$  has an effect similar to $\xi_{\rm e}$: for small 
$\xi_{\rm B}^2$ values, there are obvious breaks at the 
relativistic-Newtonian transition points; but for 
large $\xi_{\rm B}^2$ values, the break disappears, 
we could only observe a single steep line with $\alpha \geq 2.1$. 
This is also due to the dependence of $t_{\rm m}$ on $\xi_{\rm B}^2$ 
(Huang et al. 2000c). 

We have also investigated the effects of other parameters such 
as $\theta_0, n, p$ on the optical light curves (Huang et al. 2000c). 
When $\theta_0 \geq 0.4$, the light curve becomes very similar to 
that of an isotropic fireball and no steep break exists. In the 
case of a dense medium ($n \geq 10^3$ cm$^{-3}$), the expansion 
becomes non-relativistic quickly, so that the break also disappears 
and we could only observe a quick decay with $\alpha \geq 2.1$. But 
generally speaking, $p$ does not affect the break notably. 

  \begin{figure}[htb]
  \begin{center}
  \leavevmode
  \centerline{ \hbox{ \hbox{} \hspace{0.5in}
  \epsfig{figure=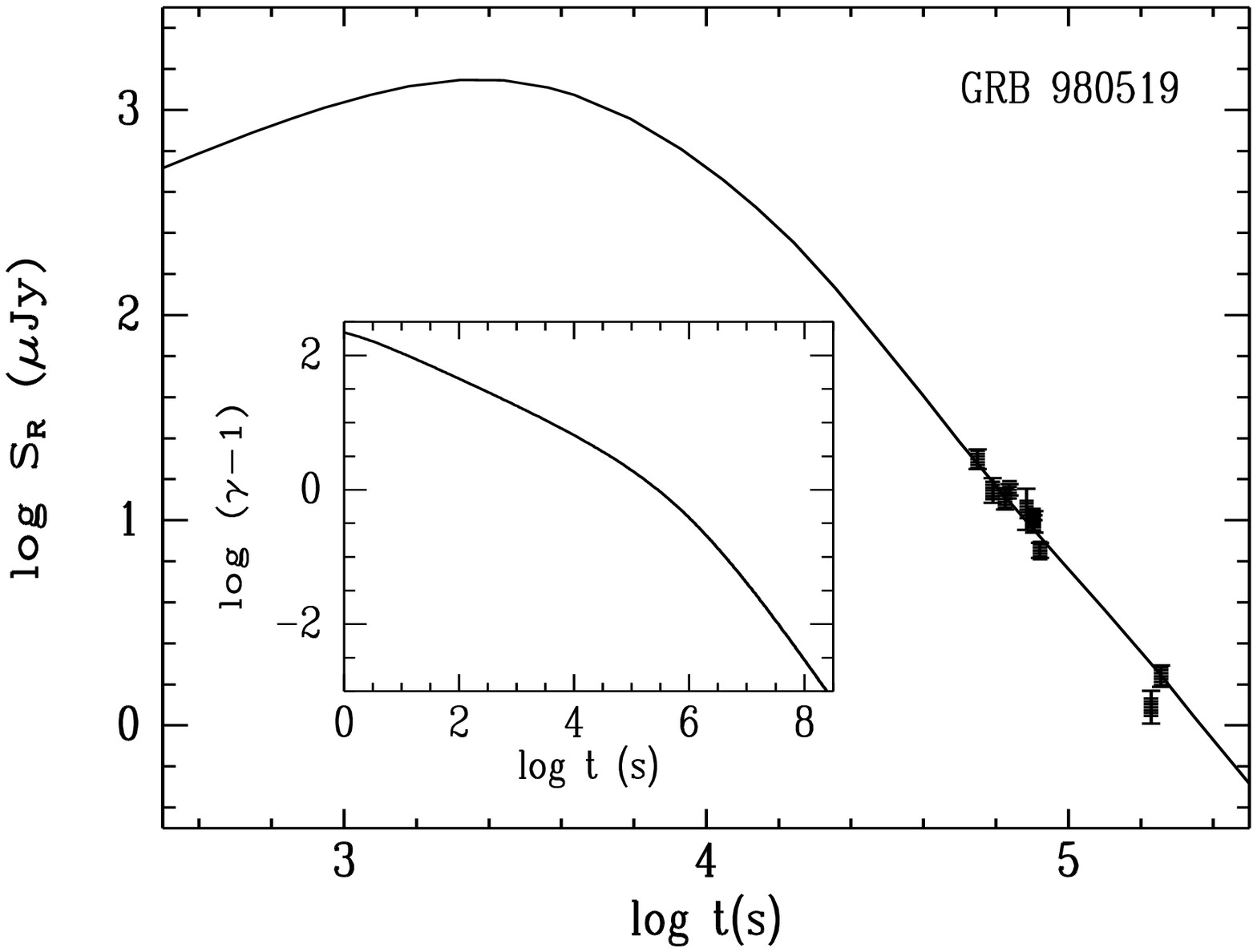,width=2.38in,height=2.0in,angle=-90,
bbllx=180pt, bblly=250pt, bburx=560pt, bbury=710pt}
  \hspace{0.5in}
  \epsfig{figure=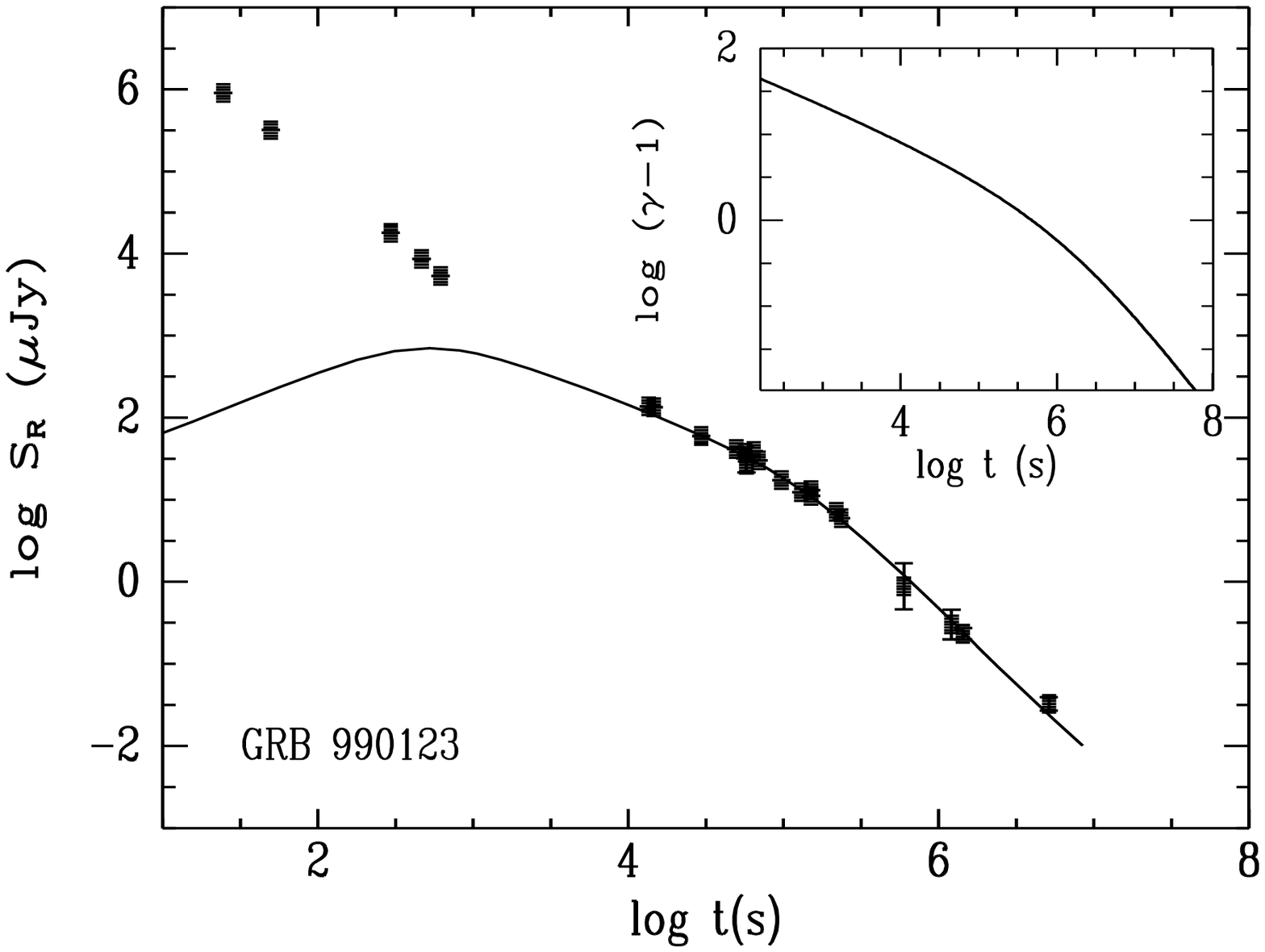,width=2.38in,height=2.0in,angle=-90,
bbllx=180pt, bblly=190pt, bburx=560pt, bbury=650pt}
  }}
  \centerline{ \hbox{ \hbox{} \hspace{0.5in}
  \epsfig{figure=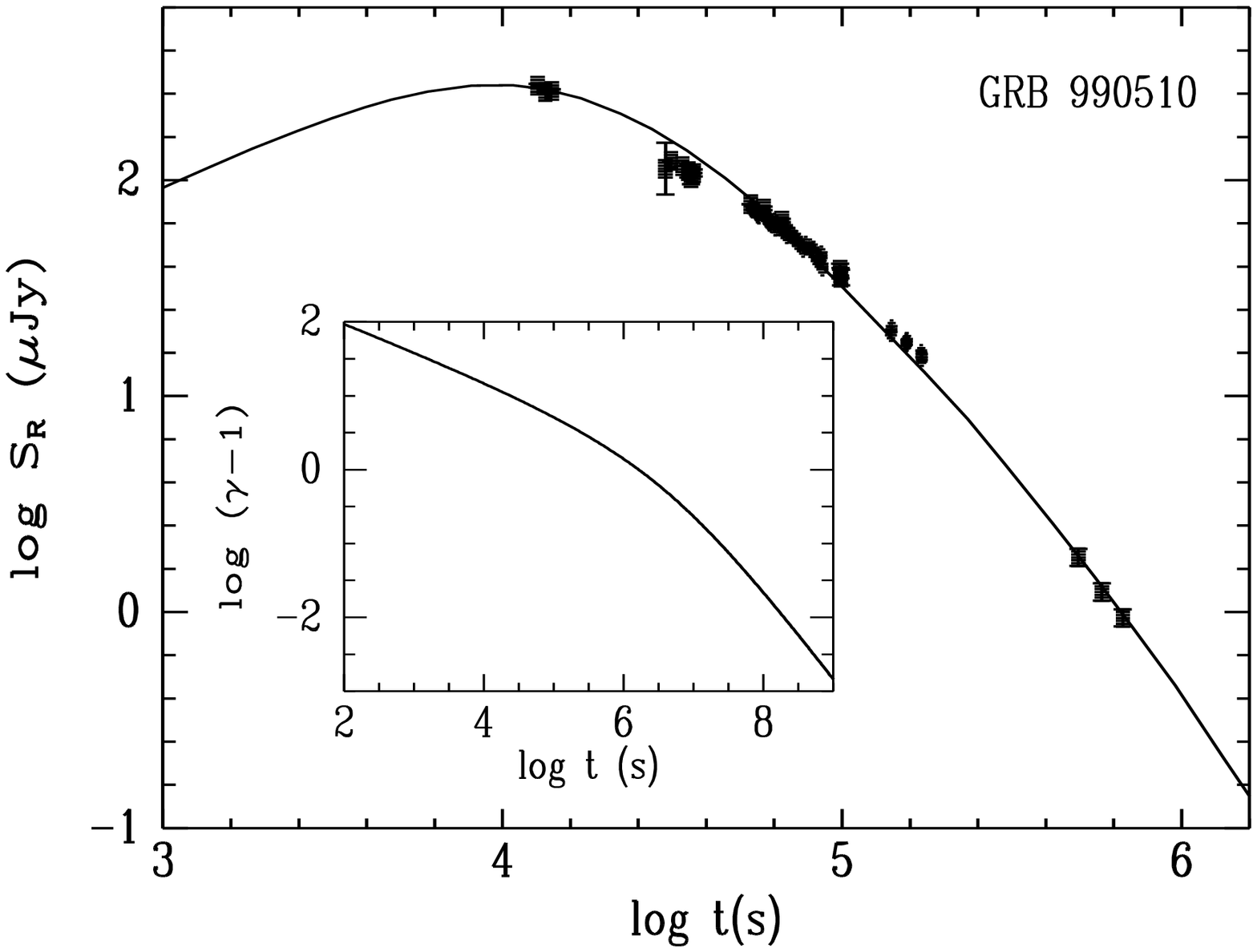,width=2.38in,height=2.0in,angle=-90,
bbllx=135pt, bblly=250pt, bburx=515pt, bbury=710pt}
  \hspace{0.5in}
  \epsfig{figure=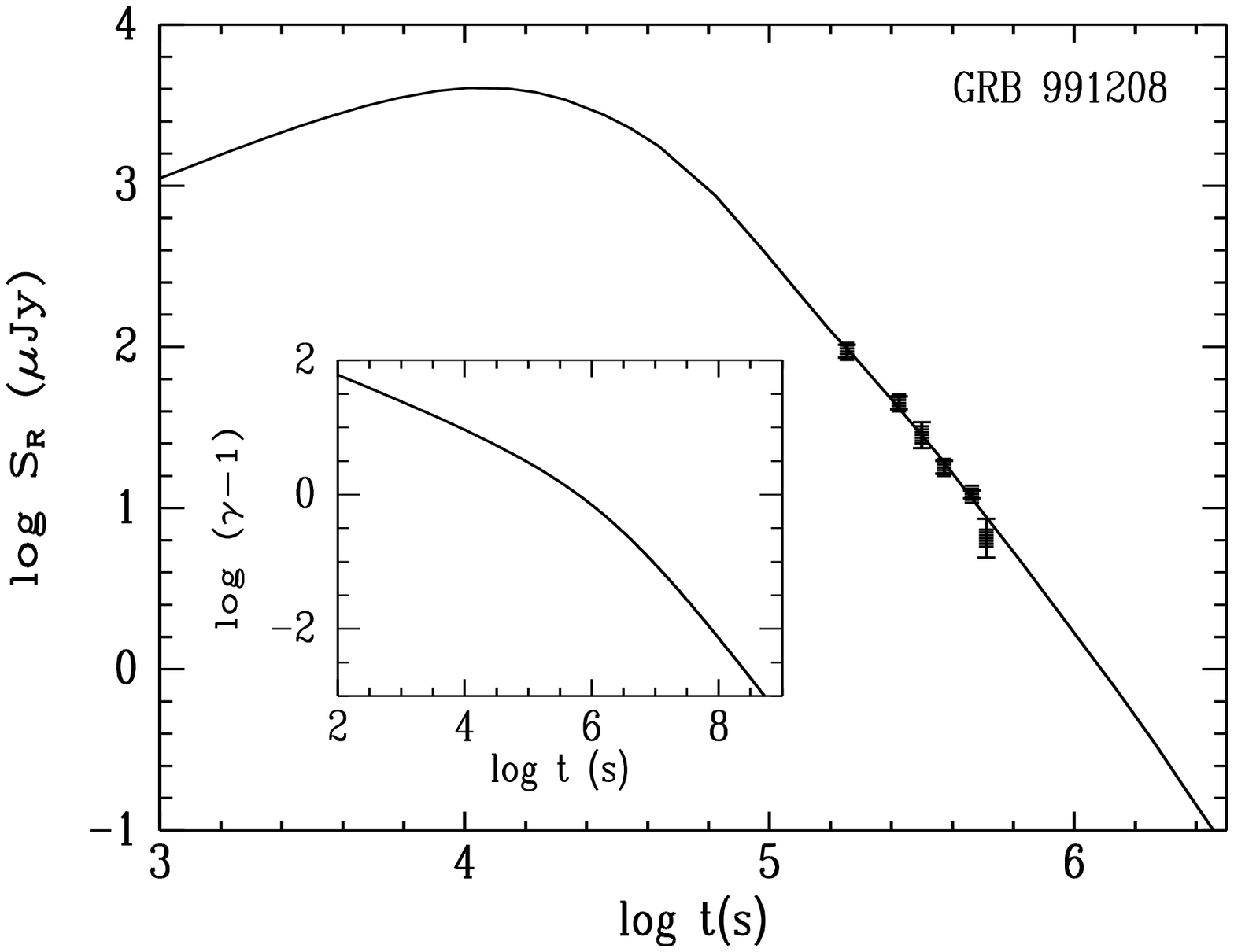,width=2.38in,height=2.0in,angle=-90,
bbllx=135pt, bblly=190pt, bburx=515pt, bbury=650pt}
  }}
\begin{minipage}{12cm}
\caption[]{Optical afterglows from GRBs 980519, 990123, 990510, 
991208 and our best fit to them by employing the refined 
jet model. Observed data are collected from the literature. Insets
show the evolution of $\gamma$ in our models (Huang et al. 2000b).}
\label{ssmsbengps}
\end{minipage}
  \end{center}
  \end{figure}

Optical afterglows from GRB 990123, 990510 are characterized by
an obvious break in the light curve, and afterglows from
GRB 970228, 980326, 980519, 991208 faded rapidly. We suggest that
these phenomena are due to beaming effects. We have fitted these
afterglows based on our refined jet model and find that the
observations can be reproduced easily with a universal initial
half opening angle $\theta_0 \sim 0.1$ (Fig. 5). 
The obvious light curve
break in GRB 990123 is due to the relativistic-Newtonian transition
of the beamed ejecta, and the rapidly fading afterglows come from
synchrotron emissions during the mildly relativistic and
non-relativistic phases. We thus strongly suggest that the rapid
fading of afterglows currently observed in some GRBs is evidence
for beaming in these cases.
 
\section{Orphan Afterglows}

The concept of orphan afterglows was first clearly proposed by Rhoads (1997), 
who pointed out that 
due to relativistic beaming effects, $\gamma$-ray radiation from the vast 
majority of jetted GRBs cannot be observed, but the corresponding late time 
afterglow emission is less beamed and can safely reach us. These afterglows 
are called orphan afterglows, which means they are not associated with any
detectable GRBs. The ratio of the orphan afterglow rate to the GRB rate 
might allow measurement of the GRB collimation angle 
(Totani \& Panaitescu 2002). Great expectations have
been put on this method (Nakar et al. 2002; Vanden Berk et al. 2002). In fact, the
absence of large numbers of orphan afterglows in many surveys has been 
regarded as evidence 
that the collimation cannot be extreme (Rhoads 1997;  
Greiner et al. 1999).

\begin{figure}[htb]
\vspace{-1.5cm}
  \begin{center}
  \leavevmode
  \centerline{ \hbox{ \hbox{} \hspace{0.5in}
  \epsfig{figure=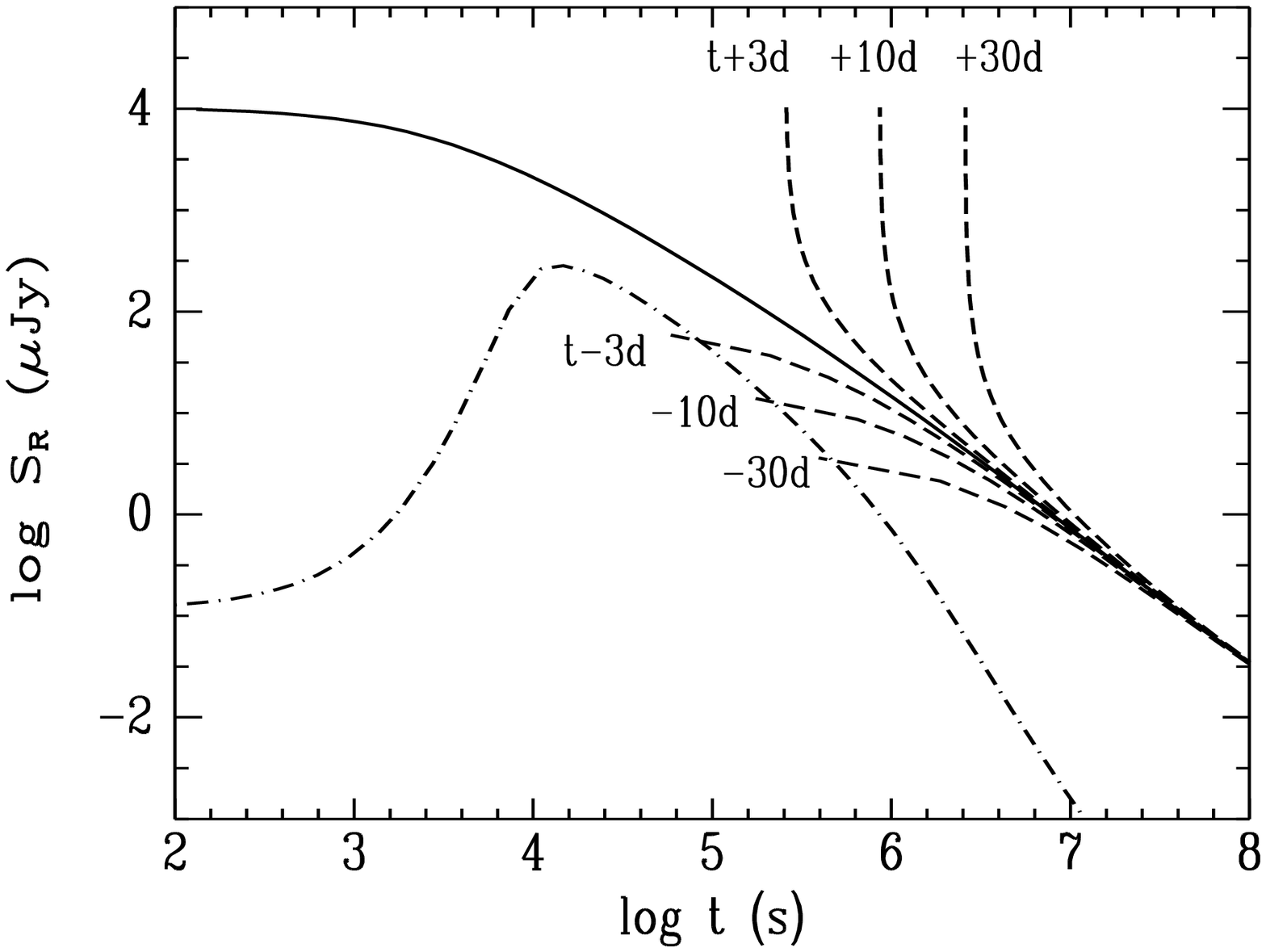,width=6.8cm,height=60mm,angle=-90,
bbllx=30pt, bblly=180pt, bburx=480pt, bbury=670pt}
  \hspace{0.5in}
  \epsfig{figure=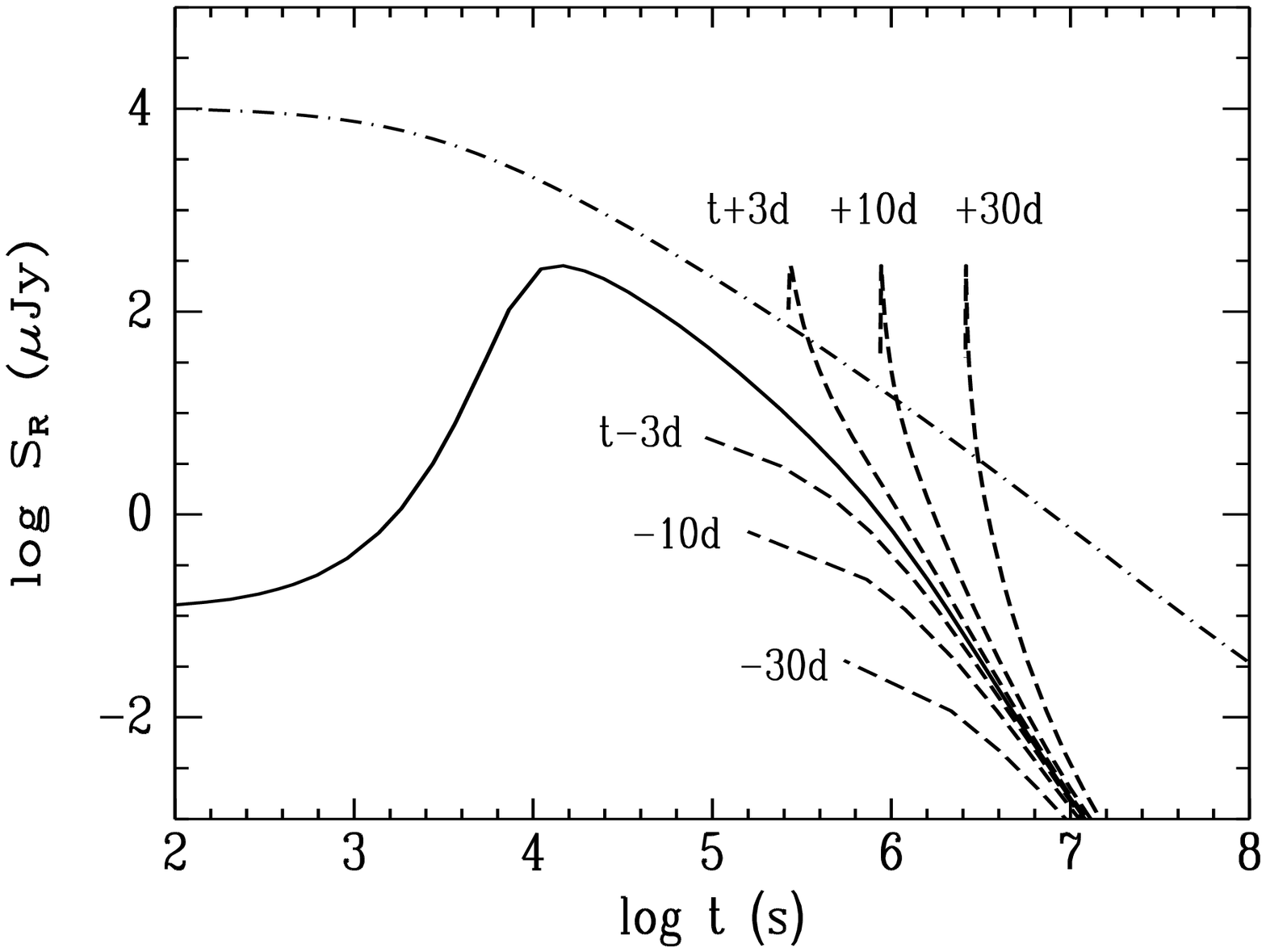,width=6.8cm,height=60mm,angle=-90,
bbllx=30pt, bblly=165pt, bburx=480pt, bbury=655pt}
  }}
  \begin{flushright}
  \parbox[t]{2.9in} { \caption {  Direct comparison of the 
  two kinds of orphan afterglows (Huang et al. 2002a). 
  The solid line represents the optical light curve of an isotropic
  FGRB and the dash-dotted line corresponds to a 
  jetted but off-axis GRB orphan. The 
  dashed lines are drawn by shifting the solid line by 
  $t \pm 3$ d, $t \pm 10$ d, and $t \pm 30$ d 
  respectively.  }} \ \hspace{.2in} \
  \parbox[t]{2.9in} { \caption { The same as in Fig. 6, 
  but this time the light curve of 
  the jetted GRB orphan is shifted (Huang et al. 2002a). }}
  \end{flushright}
  \end{center}
  \end{figure}

However, there is a difficulty associated with the method 
(Huang et al. 2002a): there should be many ``failed gamma-ray bursts 
(FGRBs)'', i.e., baryon-contaminated fireballs with initial Lorentz factor 
$\gamma_0 \ll 100$. 
BeppoSAX team has reported 
the discovery of several anomalous events named as fast X-ray
transients, X-ray rich GRBs, or even X-ray-GRBs. They resemble 
usual GRBs except that they are extremely X-ray rich (Frontera 
et al. 2000). 
Observational data on this kind of events are being accumulated
rapidly. We propose that these events are probably just FGRBs 
(Huang et al. 2002a). FGRBs cannot be observed in gamma-rays, but their
long-lasting afterglows are detectable, thus they will also manifest
themselves as orphan afterglows. 
In short, we cannot omit an important fact: if GRBs are really due 
to isotropic fireballs, then there should be much more failed GRBs. 
So, the simple discovery of orphan afterglows does not necessarily 
indicate that GRBs are beamed (Huang et al. 2002a). 

Theoretically, when orphan afterglows 
are really discovered observationally, it is still risky to conclude that 
GRBs are beamed. We should study these orphans carefully to determine 
whether they come from FGRBs or Jetted GRBs. 
Unfortunately, this is not an easy task.
The major problem is that for orphan afterglow observations, 
the derivation of a 
$\log S_{\rm \nu}$ --- $\log t$ light curve is not direct: we 
do not know the trigger time so that the exact value of $t$ for each 
observed data point cannot be determined (Huang et al. 2002a). 

In Figure 6, we compare the theoretical $\log S_{\rm \nu}$ --- $\log t$ 
light curves of optical afterglows 
from FGRBs and jetted but off-axis GRBs directly. To investigate the 
influence of the uncertainty in trigger time, we also shift the 
light curve of FGRBs by $t \pm 3$ d, $t \pm 10$ d and $t \pm 30$ d 
intentionally. From the dashed curves, we can see that the shape of 
the FGRB afterglow light curve is seriously affected by the uncertainty 
of the trigger time. But fortunately, these dashed curves still 
differ from the theoretical light curve of the jetted GRB orphan 
markedly, i.e., they are much flatter at very late stages. This means
it is still possible for us to discriminate them. 
In Figure~7, similar results to Figure~6 are given, but this 
time the light curve of the jetted GRB orphan is shifted. Again we see
that the two kinds of orphans can be discriminated by their late time 
behaviour. 

Figures 6 and 7 explain what we should do when an orphan afterglow is 
discovered. First, we have to assume a trigger time for it arbitrarily, 
so that the logarithmic 
light curve can be plotted. We then need to change the 
trigger time to many other values to see how the light curve is 
affected. In all our plots, we should pay special attention to the 
late time behaviour, which will be less affected by the uncertainty in 
the trigger time. If the slope tends to be $\sim -1.0$ --- $-1.3$, then
the orphan afterglow may come from an FGRB event. But if the slope 
tends to be steeper than $\sim -2.0$, then it is very likely from 
a jetted but off-axis GRB (Huang et al. 2002a). 

However, we must bear in mind that it is in fact not an easy 
task. First, to take the process we need 
to follow the orphan as long as possible, and 
the simple discovery of an orphan is obviously insufficient. 
Note that currently optical afterglows from most 
well-localized GRBs can be observed for only less than 100 days. 
It is quite unlikely that we can follow an orphan for 
a period longer than that.  
Second, since the orphan is usually very faint, 
errors in the measured 
magnitudes will seriously prevent us from deriving the slope.  
Due to all these difficulties, a satisfactory light curve
is usually hard to get for most orphans (Huang et al. 2002a). 
We see that measurement of the GRB 
beaming angle using orphan searches is not as simple as we 
originally expected. In fact, it is impractical to some extent
(Huang et al. 2002a).  

Anyway, there are
still some other possible solutions that may help to determine 
the onset of an orphan afterglow (Huang et al. 2002a).  
Firstly, of course we should improve our detection limit so 
that the orphan afterglow could be followed as long as possible.
The longer we observe, the more likely that we can get the true
late-time light curve slope. Secondly, we know that FGRBs usually 
manifest themselves as fast X-ray transients or X-ray-GRBs. If 
an orphan can be identified to associate with such a transient, 
then it is most likely an FGRB one. In this case, the trigger 
time can be well determined.  
Thirdly, maybe in some rare cases we are 
so lucky that the rising phase of the orphan could be observed.
For a jetted GRB orphan the maximum optical flux is usually 
reached within one or two days and for an FGRB orphan it is even
within hours. Then the uncertainty in trigger time is greatly 
reduced. Additionally, a jetted GRB orphan differs markedly 
from an FGRB one during the rising phase. The former can be 
brightened by more than one magnitude in several hours (see 
Figures~6 --- 7), while the brightening of the latter can hardly 
be observed. So, if an orphan afterglow with a short period of 
brightening is observed, then it is most likely a jetted GRB 
orphan. Of course, we should first be certain that it is not 
a supernova. 

Fourthly, valuable clues may come from radio observations. In radio
bands, the light curve should be highly variable at early stages 
due to interstellar medium scintillation, and it will become much
smoother at late times. So the variability in radio light curves 
provides useful information on the trigger time. And fifthly, in 
the future maybe gravitational wave radiation 
or neutrino radiation associated with GRBs
could be detected due to progresses in technology, then the trigger
time of an orphan could be determined directly and accurately.
Sixthly, the microlensing 
effect may be of some help. Since the size of the radiation zone
of a jetted GRB orphan is much smaller than that of an FGRB one, 
they should behave differently when microlensed (Huang et al. 2002a). 

Finally, although a successful detection of some orphan afterglows
does not directly mean that GRBs be collimated, the negative 
detection of any orphans can always place both a stringent lower limit
on the beaming angle for GRBs and a reasonable upper limit for the 
rate of FGRBs (Huang et al. 2002a).  

\begin{figure}[htb]
\vspace{-1.5cm}
  \begin{center}
  \leavevmode
  \centerline{ \hbox{ \hbox{} \hspace{0.5in}
  \epsfig{figure=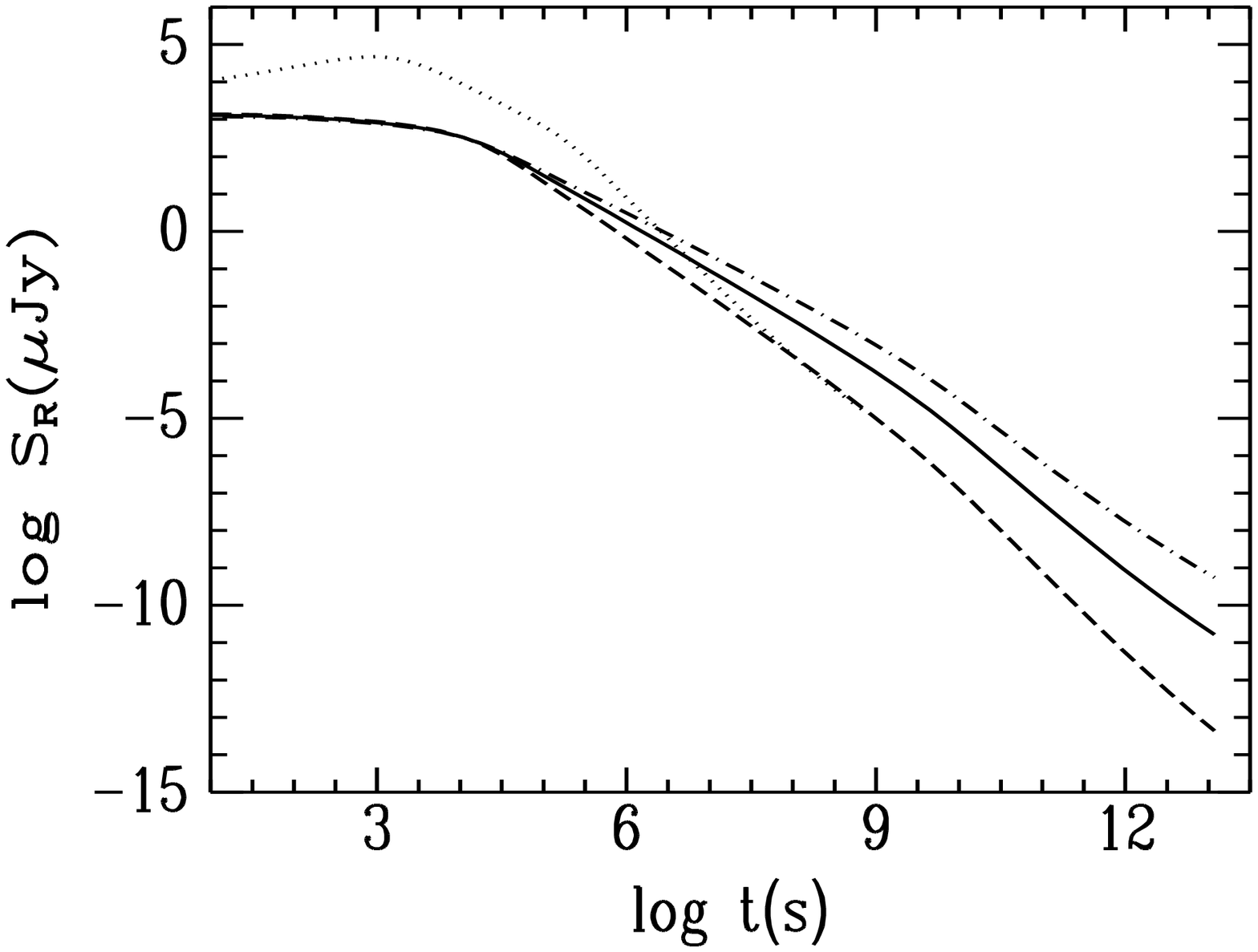,width=6.8cm,height=60mm,angle=-90,
bbllx=30pt, bblly=180pt, bburx=480pt, bbury=670pt}
  \hspace{0.5in}
  \epsfig{figure=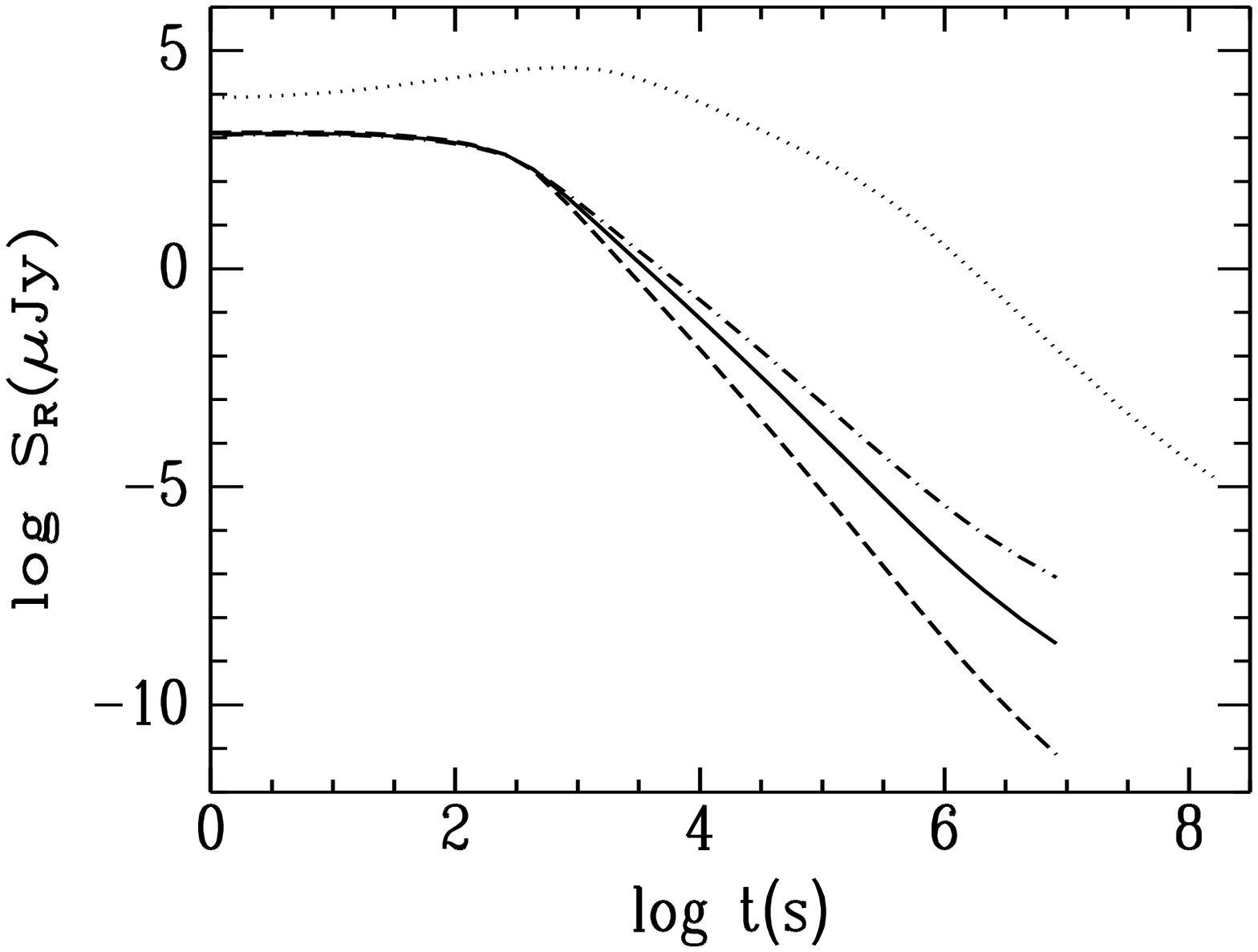,width=6.8cm,height=60mm,angle=-90,
bbllx=30pt, bblly=165pt, bburx=480pt, bbury=655pt}
  }}
  \begin{flushright}
  \parbox[t]{2.9in} { \caption { R band afterglows from beamed GRB 
  ejecta without lateral expansion
  ($v_{\perp} \equiv 0$). The dotted line corresponds to a
  conical jet with $p = 2.5$, other lines are for cylindrical jets which
  differ only in the parameter $p$. The dash-dotted, solid and the dashed 
  line corresponds to $p = 2.2$, 2.5 and 3.0 respectively. 
  The breaks at $t \sim 10^9$ s in the 
  light curves for cylindrical jets are due to cooling of electrons
  (Cheng et al. 2001).  }} \ \hspace{.2in} \
  \parbox[t]{2.9in} { \caption { R band afterglows from beamed GRB ejecta 
  with lateral expansion
  ($v_{\perp} \equiv c_{\rm s}$). The dotted line corresponds to a
  conical jet with $p = 2.5$, other lines are for cylindrical jets which
  differ only in the parameter $p$. The dash-dotted, solid and the dashed 
  line corresponds to $p = 2.2$, 2.5 and 3.0 respectively. 
  Note that the cylindrical jets here are already 
  non-relativistic when $t > 10^6$ s (Cheng et al. 2001). }}
  \end{flushright}
  \end{center}
  \end{figure}

\section{Cylindrical Jets}

Nearly all previous discussions on beaming effects in gamma-ray bursts
have assumed a conical geometry. However, more and more observations on
relativistic jets in radio galaxies, active galactic nuclei, and
``microquasars'' in the Galaxy have shown that many of these outflows
are not conical, but cylindrical, i.e., they maintain constant cross
sections at large scales.
Thus it is necessary to discuss the possibility
that gamma-ray bursts may be due to highly collimated cylindrical jets,
not conical ones. In fact, this idea has already been suggested
as GRB trigger mechanism by Dar (1998).

Dynamical evolution of cylindrical jets
and their afterglows have been discussed in great detail by Cheng et al.
(2001). Both analytical and numerical results are
presented. It is shown that when the lateral expansion is not taken into
account, a cylindrical jet typically remains to be highly relativistic 
for $\sim 10^8$ --- $10^9$ s. During this relativistic phase, the optical 
afterglow decays as $S_{\nu} \propto t^{-p/2}$ at first. 
Then the light curve steepens to be $S_{\nu} \propto t^{-(p+1)/2}$ due to 
cooling of electrons. After entering the non-relativistic phase (i.e., 
$t \geq 10^{11}$ s), the afterglow is $S_{\nu} \propto t^{-(5p-4)/6}$.
But if the cylindrical jet expands laterally at co-moving sound speed, 
then the decay becomes $S_{\nu} \propto t^{-p}$ 
and $S_{\nu} \propto t^{-(15p-21)/10}$ --- $t^{-(15p-20)/10}$ in the
ultra-relativistic and non-relativistic phase respectively. Note that
in both cases, the light curve turns flatter after the
relativistic-Newtonian transition point, which differs markedly from
the behaviour of a conical jet. 

It is suggested that some gamma-ray
bursts with afterglows decaying as $t^{-1.1}$ --- $t^{-1.3}$ may be
due to cylindrical jets, not necessarily isotropic fireballs
(Huang et al. 2002b).

\section{Concluding Remarks}

We investigate beaming effects in GRBs numerically. It is found 
that the light curve break of optical afterglows is not obvious 
within the relativistic phase. But an obvious break can truly be 
found at the relativistic-Newtonian transition point. Further 
investigations show that the break is parameter dependent, 
but afterglows from a jet in the non-relativistic phase 
are uniformly characterized by a quick decay (such as $t^{-2}$). 
It is also shown that measure of GRB beaming by using orphan afterglow
surveys is not as easy as original expectancy. 

As for the geometry of GRB beaming, we suggest that the jet could also
be cylindrical. Afterglows from cylindrical jets with or without
lateral expansion have been discussed. It is shown that those 
GRBs with a flat afterglow light curve could be well accounted for 
by a cylindrical jet without lateral expansion. 

The most difficult enigma in GRBs is the progenitor. Beaming effects
can provide important clues. Other helpful hints may come from 
X-ray observations (Antonelli et al. 2000).

\vspace{1cm}
This work was supported by The Foundation for the Author of
National Excellent Doctoral Dissertation of P. R. China (Project
No: 200125), the Special Funds for Major State Basic Research
Projects, the National Natural Science Foundation of China, and
the National 973 Project (NKBRSF G19990754).

\clearpage

\centerline{\Large\bf Reference}

\noindent
Antonelli L.A. {\em et al.}, 2000, ApJ, 545, L39 \\
Castro-Tirado A. {\em et al.}, 1999, Sci, 283, 2069 \\
Cheng K.S., Huang Y.F., Lu T., 2001, MNRAS, 325, 599 (astro-ph/0102463) \\
Dai Z.G. \& Gou L.J., 2001, ApJ, 552, 72 \\
Dai Z.G. \& Cheng K.S., 2001, ApJ, 558, L109 \\
Dar A., 1998, ApJ, 500, L93 \\
Frontera F. et al., 2000, ApJ, 540, 697  \\
Gou L.J. {\em et al.}, 2001, A\&A, 368, 464 \\
Greiner J., Voges W., Boller T., Hartmann D., 1999, A\&AS, 138, 441 \\
Halpern J.P. {\em et al.}, 2000, ApJ, 543, 697  \\
Huang Y.F., 2000, astro-ph/0008177 \\
Huang Y.F., Dai Z.G. \& Lu T., 1998a, A\&A, 336, L69 \\
Huang Y.F., Dai Z.G. \& Lu T., 1998b, MNRAS, 298, 459  \\
Huang Y.F., Dai Z.G. \& T. Lu, 1999, MNRAS, 309, 513. \\
Huang Y.F. {\em et al.}, 2000a, ApJ, 543, 90 (astro-ph/9910493) \\
Huang Y.F., Dai Z.G. \& Lu T., 2000b, A\&A, 355, L43 (astro-ph/0002433) \\
Huang Y.F., Dai Z.G. \& Lu T., 2000c, MNRAS, 316, 943 (astro-ph/0005549) \\
Huang Y.F., Dai Z.G., Lu T., 2002a, MNRAS, 332, 735 (astro-ph/0112469) \\
Huang Y.F. {\em et al.}, 2002b, Chinese Astron. Astrophys., accepted \\
Kulkarni S.R. {\em et al.}, 1999, Nat, 398, 389 \\
M\'{e}sz\'{a}ros P. \& Rees M.J., 1999, MNRAS, 306, L39 \\
Nakar E., Piran T., Granot J., 2002, ApJ, accepted (astro-ph/0204203) \\
Panaitescu A. \& M\'{e}sz\'{a}ros P., 1999, ApJ, 526, 707 \\
Ramirez-Ruiz E., Lloyd-Ronning N.M., 2002, New Astronomy, accepted (astro-ph/0203447) \\
Rhoads J., 1997, ApJ, 487, L1 \\
Totani T., Panaitescu A., 2002, ApJ, accepted (astro-ph/0204258) \\
van Paradijs J., Kouveliotou C. \& Wijers R.A.M.J., 2000, ARA\&A, 38, 379 \\
Vanden Berk D.E. {\em et al.}, 2002, ApJ, in press (astro-ph/0111054) \\
Zhang B., M\'{e}sz\'{a}ros P., 2002, ApJ, 571, 876 \\

\end{document}